\documentclass[%
reprint,
amsmath,amssymb,
prb,
]{revtex4-1}

\usepackage{graphicx}
\usepackage{dcolumn}
\usepackage{bm}
\usepackage{color}
\usepackage{xfrac}
\usepackage[font=small,labelfont=bf,justification=justified,format=plain]{caption}
\usepackage{subcaption}

\begin{document}

\preprint{APS/123-QED}

\title{Efficient technique for computational design of thermoelectric materials}

\author{Maribel N\'u\~nez-Valdez}
\email{nunez025@umn.edu}
\affiliation{Moscow Institute of Physics and Technology, Dolgoprudny, Moscow Region, Russia}

\author{Zahed Allahyari}
\affiliation{
  Skolkovo Institute of Science and Technology, Skolkovo Innovation Center, 3 Nobel St., Moscow 143026, Russia \\
  Moscow Institute of Physics and Technology, Dolgoprudny, Moscow Region, Russia
}

\author{\textcolor{black}{Tao Fan}}
\affiliation{
  \textcolor{black}{Northwestern Polytechnical University, School of Materials Science and Engineering,
Xi’an 710072, China}
}

\author{Artem R. Oganov}
\affiliation{
Skolkovo Institute of Science and Technology, Skolkovo Innovation Center, 3 Nobel St., Moscow 143026, Russia\\
Moscow Institute of Physics and Technology, Dolgoprudny, Moscow Region, Russia\\
International Center for Materials Design, Northwestern Polytechnical University, Xi'an,710072, China
}%

\date{\today}
\begin{abstract}
Efficient thermoelectric materials are highly desirable, and the quest for finding them has intensified as they could be promising alternatives to fossil energy sources. Here we present a general first-principles approach to predict, in multicomponent systems, efficient thermoelectric compounds. The method combines a robust evolutionary algorithm, a Pareto multiobjective optimization, density functional theory and a Boltzmann semi-classical calculation of thermoelectric efficiency. To test the performance and reliability of our overall framework, we use the well-known system Bi$_2$Te$_3$-Sb$_2$Te$_3$.  
\end{abstract}
\pacs{Valid PACS appear here}
\maketitle
\section{Introduction}
Finding alternatives to fossil energy sources is a priority for many scientific and engineering fields. Thermoelectric energy conversion, that is converting waste heat into electricity, is a particularly attractive method as thermoelectric devices are highly reliable, integrable, stable, and compact\cite{HamidElsheikh2014}. Applications of thermoelectrics include conventional coolers, laser cooling, cryogenic infrared night vision equipment, telecom lasers, electronic cooling and even devices for outer space exploration\cite{pichanusakorn2010}. \\

Thus, thermoelectric materials have been intensively studied during the last decades, however the energy conversion efficiencies obtained have been quite low, thus limiting the use of thermoelectric devices as promising alternative energy sources (for a review see Ref.~[\onlinecite{Snyder2008}] and references therein).\\

The difficulties found in enhancing the thermoelectric efficiency are manifold and depend on the optimization of several, often clashing, parameters. The efficiency or figure of merit, $ZT$, that characterizes each material is given by the combination of different transport coefficients as follows:
\begin{equation}
  ZT=\frac{\sigma S^2 T}{\kappa_e+\kappa_l}
\end{equation}
where $\sigma$ is the electrical conductivity, $S$ is the Seebeck coefficient, $T$ is temperature, and $\kappa_e$ and $\kappa_l$ are electronic and lattice thermal conductivities, respectively. These quantities have been studied individually, from the experimental and theoretical points of view, and they are frequently tailored according to a particular application.  For instance, a large Seebeck coefficient is usually obtained using only one type of carriers (n-type or p-type), however materials with large $S$ tend to have low electrical conductivity $\sigma$, thus adjustments have to be made in order to maximize the figure of merit\cite{Snyder2008}. Also, great efforts have gone into minimizing the thermal conductivity, but this task is far from easy. On one hand, from the first-principles point of view, the description of electrons and holes transporting heat, $\kappa_e$, is to some extent tractable within the Boltzmann semi-classical theory and a constant relaxation time\cite{boltztrap}. On the other hand, $\kappa_l$ depends on the structure, rigidity, atomic masses and other characteristics of the lattice\cite{dmitriev2010}. 
Computations of vibrational properties and lattice thermal transport\cite{Li2014} of individual materials have been done (for a review, see e.g. Ref.~[\onlinecite{yan2015}]), but they are computationally expensive.\\

In this work, as our main contribution, we demonstrate that an evolutionary algorithm, in combination with density functional theory (DFT)\cite{dft1,dft2} and Boltzmann semi-classical calculation of transport properties can be used to find Pareto-optimal solutions in terms of energy and thermoelectric efficiency within finite time, provided a few criteria are met. The ability to predict the most stable crystal structures purely from first-principles using an evolutionary approach such as USPEX\cite{oganov_glass_2006,Oganov2008,lyakhov2013}, by providing only the chemical composition, has had a number of successful results widely discussed in the literature. Now we extend this method, to look for structures that are not only the most stable, but at the same time, possess a large figure of merit $ZT$. This type of optimization task is part of the so-called  multiobjective optimization problems. In general, there is no unique solution that can optimize all the objectives simultaneously, instead, a set of Pareto optimal solutions exist for such a problem\cite{miett99}. To accomplish such quest, careful algorithm design is very important in finding a Pareto set\cite{Deb:2001:MOU:559152}.\\

Among state-of-the-art approaches, to the best of our knowledge, the endeavor described above has not been tackled. Recently, there have been attempts of hybrid approaches, that is, methods merging in a database frame and web-based tools, first-principles calculations and experimental information with the purpose to resolve more efficient thermoelectric materials\cite{gorai2015,yan2015}. Yet, a purely {\it ab initio} method has remained elusive until now. We hope that the first-principles approach that we develop here will help and be a step further into the discovery of new efficient thermoelectric materials.\\

The remainder of this paper is organized as follows: in Section II, computational details of our first-principles method and calculations are given. In Section III, we discuss our results and implications. Finally in Section IV, a brief summary and our conclusions are outlined. \\

\begin{figure}[h]
\includegraphics[width=0.5\textwidth]{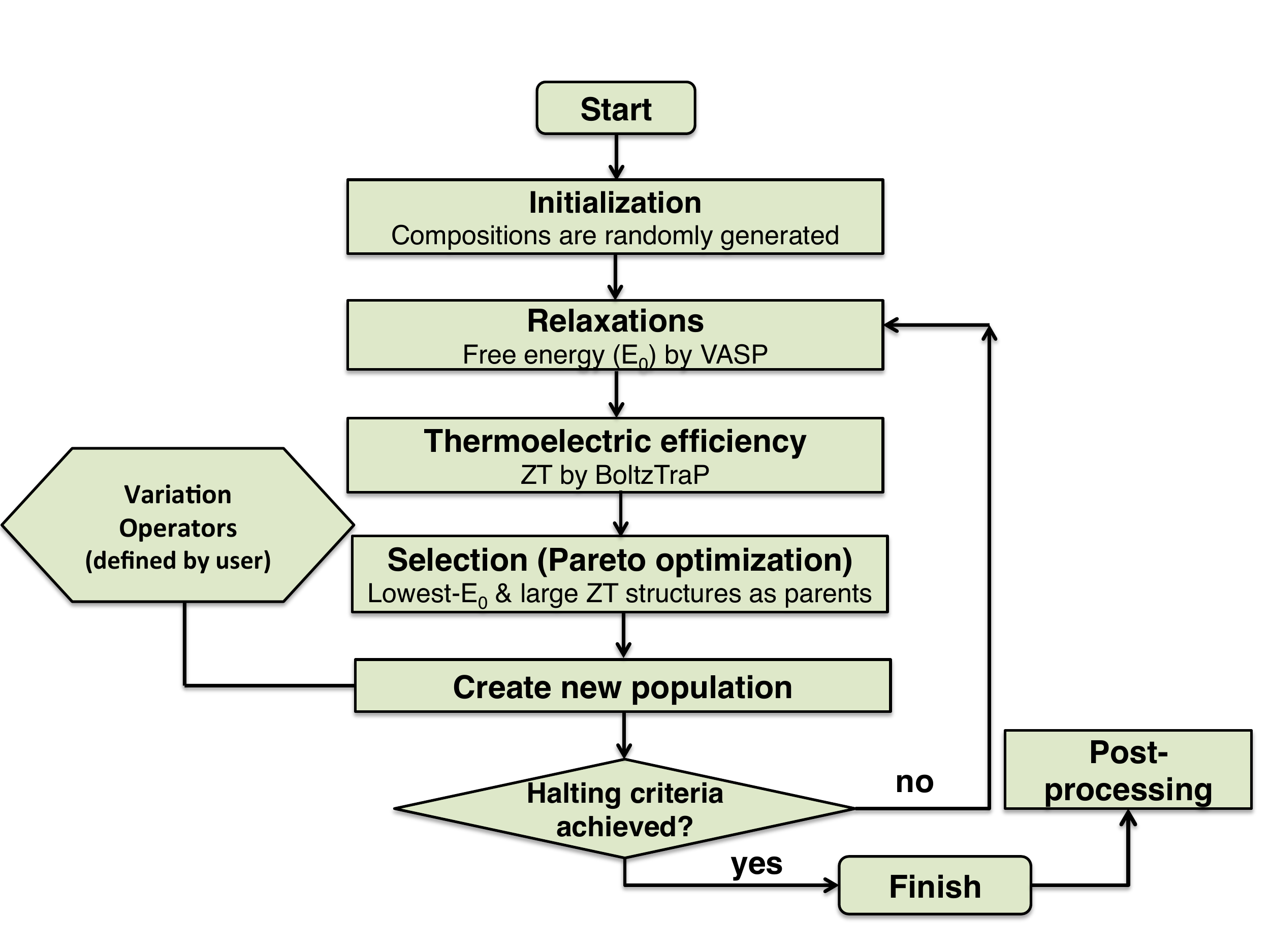}
\caption{\label{fig1} (Color online) Workflow for a variable-composition evolutionary search of efficient thermoelectric compounds using the USPEX\cite{oganov_glass_2006}, VASP\cite{kresse96a} and BoltzTraP\cite{boltztrap} codes.}
\end{figure}
\section{Computational Details and Methods}
Calculations presented here were performed using the variable-composition evolutionary algorithm USPEX\cite{oganov_glass_2006,oganov_lyakhov_valle_2011}, interfaced with VASP\cite{kresse96a} and BoltzTraP\cite{boltztrap}. The Pareto implementation simultaneously monitored the energy and thermoelectric efficiency of the structures and constructed Pareto fronts as follows.\\

In general, from a starting set of structures $\{S_n\}$ ($n=1,\ldots,N$), and with each structure having properties $P_m$ ($m=1,\ldots,M$), we say that structure $S_i$ Pareto dominates structure $S_j$ ($i\ne j$) in terms of properties $P_k$ and $P_l$ ($k\ne l$) if:
\begin{eqnarray}
  P_k(S_i)&< &P_k(S_j), \nonumber\\
  \mbox{and}& &\nonumber\\
P_l(S_i)&\leq& P_l(S_j). \nonumber
\end{eqnarray}
A structure $S^*$ not dominated by any other is, therefore, Pareto optimal. The subset of all Pareto optimal structures $\{S_t\}$ ($t=1,\ldots,T$ and $T<N$) constitutes the Pareto front 1. These $S_t$ structures are then removed from the initial set of $N$ individuals, and the cycle is repeated to find successive Pareto fronts 2, 3,$\ldots$ until all $N$ structures are classified. \\

These Pareto fronts are very valuable since they provide one way to ``make partial decisions'' in multivalue problems. The ``front 1'' contains the most optimal solutions and delimits the rest of them in the sense that, for every member on front 1 an improvement of one of its properties is not possible without degrading the other. In our approach, the two properties $P_{1,2}$ to be optimized {\it simultaneously} are the energy and figure of merit. Here, we stress that the optimization involves both properties to achieve success since optimizing only, e.g., $ZT$ would probably lead to very efficient materials but energetically unstable. Thus, the first generation of structures was produced by USPEX randomly (up to 18 atoms per primitive cell), while successive generations were created by applying variation operators (genetic mutation 20\%, random symmetric generator 20\%, soft mutation 20\%, transmutation 20\%, lattice mutation 10\% and random topology 10\%). 
The initial population consisted of 150 structures and subsequent generations were created with 90 structures. Each individual underwent a series of relaxations (five in total and from lower to higher precision), the thermoelectric efficiency $ZT$ was calculated next, and the fittest individuals, that is, the structures with the lowest energies and largest efficiencies were selected as parents to produce a new generation using the variation operators described above. We observe that with our algorithm, we find and optimize the maximum of the thermoelectric efficiencies by `minimizing the negative of $ZT$'. Each new generation reentered the cycle, which stopped after 40 generations were computed. Interesting structures produced by USPEX were re-checked using higher precision for structural, electronic and thermoelectric efficiency in the post-processing (see Fig. 1). Another technical aspect is that we disregarded spin-orbit coupling in the evolutionary search as its account would be too costly, however, overlooking it seemed not to have interfered with the general outcome of the simulation.\\

All structures were relaxed using density functional calculations within the projector-augmented plane wave (PAW)\cite{blochl,kresse99} method. We used the general-gradient approximation (GGA) in the Perdew-Burke-Ernzerhof (PBE)\cite{perdew2} prescription for the exchange-correlation potential. We also employed the default PAW potentials with the valence configurations 5$d^{10}$6$s^2$6$p^3$ for Bi, 5$s^2$5$p^3$ for Sb, and 5$s^2$5$p^4$ for Te. For structure relaxations, we used plane wave cutoff of 500 eV. Transport properties' calculations for the figure of merit $ZT$ (Eq. 1), at a given temperature (here we used 300 K) were carried out based on Boltzmann transport theory within the constant scattering-time approximation using the BoltzTraP code\cite{boltztrap}. We notice that within this framework, the lattice thermal conductivity $\kappa_l$ is neglected, since its consideration would be prohibitively expensive for high-throughput searches. \\

\section{Results}
\begin{figure}[h]
\includegraphics[width=0.5\textwidth]{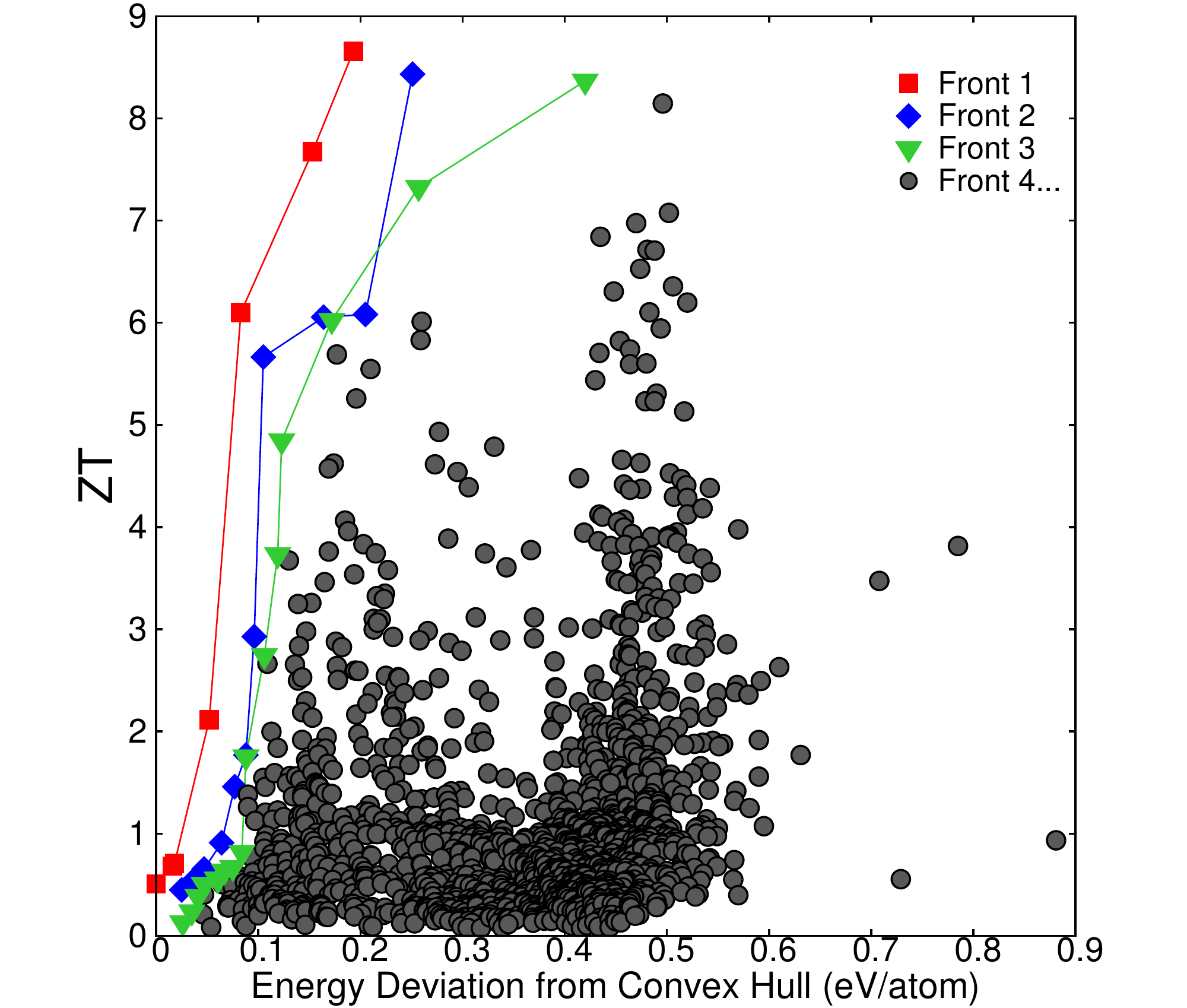}
\caption{\label{fig2} (Color online) Raw classification of structures (all symbols) according to stability and thermoelectric efficiency ($ZT$) at 300 K. The first three Pareto fronts are highlighted with ``Front 1'', representing the best Pareto optimal choices of stable and metastable thermoelectric efficient compounds.} 
\end{figure}
\begin{figure}[h]
\begin{subfigure}{0.21\textwidth}
  \includegraphics[width=\textwidth]{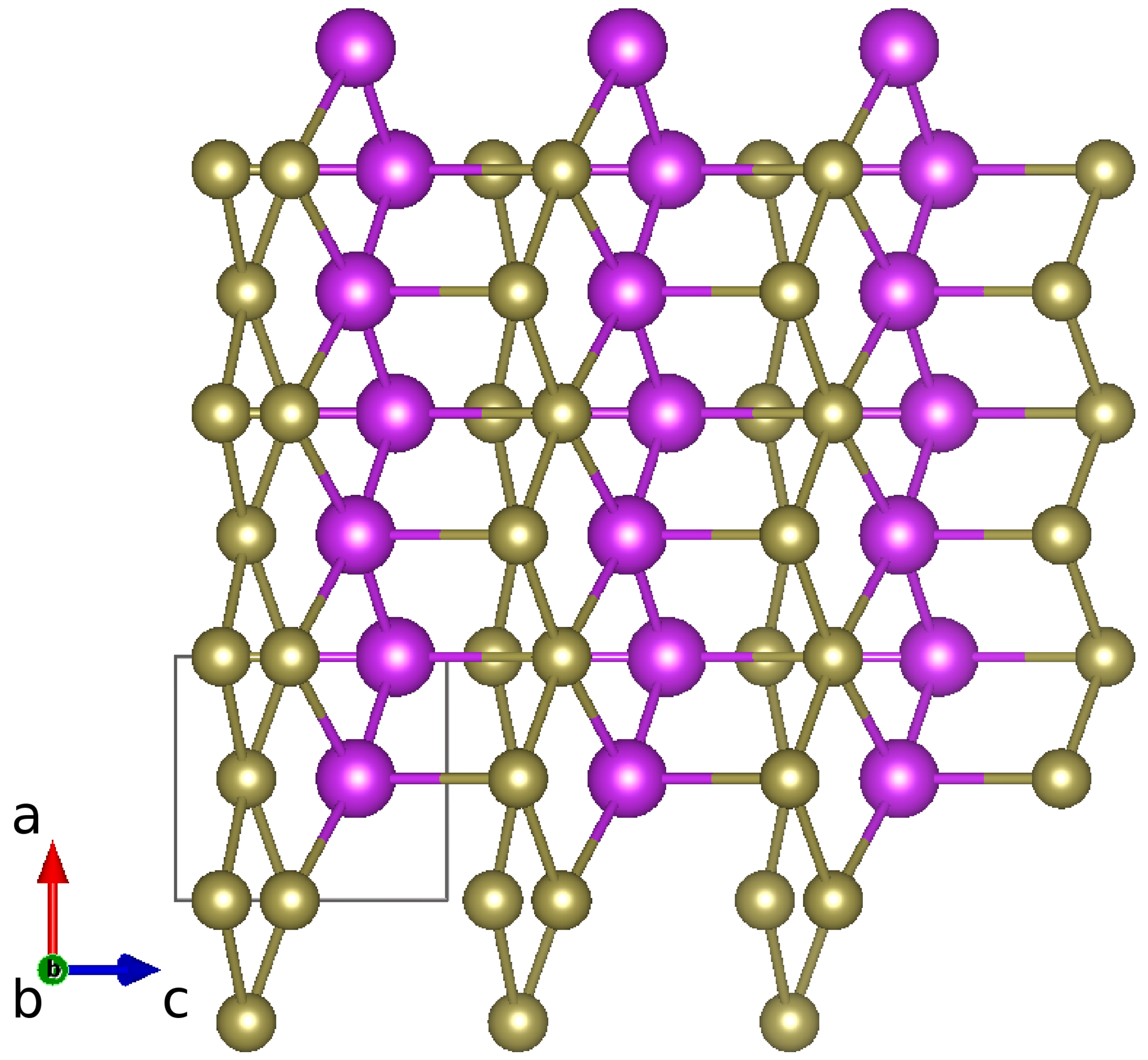}
  \caption{Monoclinic symmetry ($Pm$)}
  \label{metastable_a}
\end{subfigure}
\begin{subfigure}{0.237\textwidth}
        \includegraphics[width=\textwidth]{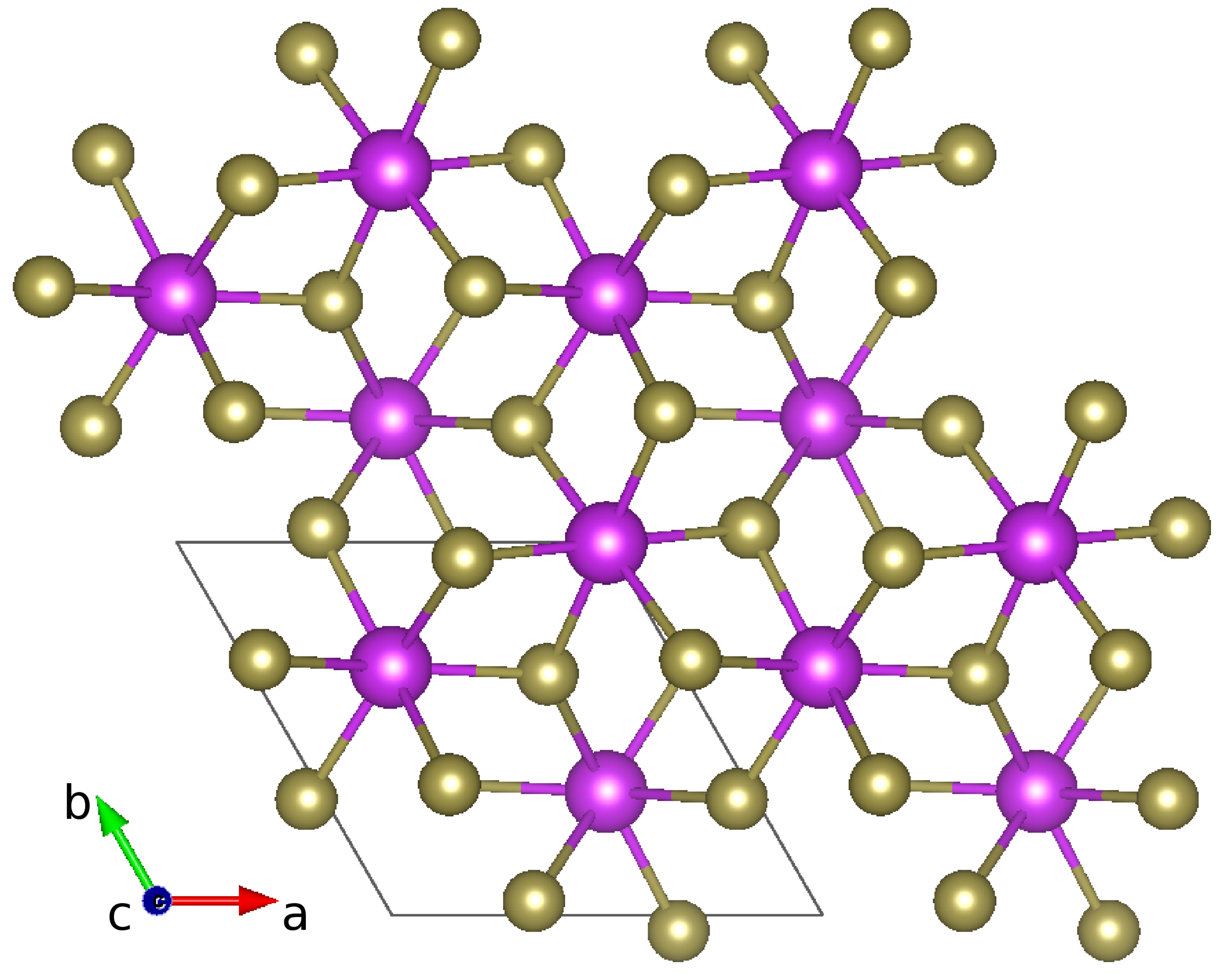}
        \caption{Hexagonal symmetry ($P6_3$)}
        \label{metastable_b}
    \end{subfigure}
\caption{(Color online) Examples of metastable structures belonging to the Pareto Front 1. (All crystal structures in this paper are visualized using VESTA\cite{vesta}.)}
\label{metastable_ab}
\end{figure}
To validate our method, we looked into structures with Bi, Sb and Te as desired atom types in them.  Bi$_2$Te$_3$, Sb$_2$Te$_3$ and their solid solutions  are some of the most studied bulk materials since they are well established as outstanding thermoelectric materials at room temperature (300 K) and ambient pressure. Bi$_2$Te$_3$ and Sb$_2$Te$_3$ are isostructural and under ambient conditions crystallize in a rhombohedral structure (space group No. 166, $R\overline{3}m$ and $Z=3$)\cite{database}. Thus, the Bi-Sb-Te system provided us with ideal ingredients for testing our modeling.\\ 

Our evolutionary multiobjective optimization successfully ran and produced 4829 structures that were classified and ranked in Pareto fronts. In Fig. 2 we show the total of the structures, with the first three Pareto fronts highlighted as examples, \textcolor{black}{in the $ZT$-{\it `Thermodynamic stability'} space. This stability was quantified for each structure by calculating its energy deviation from the thermodynamic {\it convex hull}, which is determined by the energy of formation, $E_f$. For instance, for an ``A-B'' system, $E_f(A_xB_y)=E(A_xB_y)-xE(A)-yE(B)$, and thus any thermodinamically stable phase against decomposition into other binaries or the elements, is located on the convex hull (when the pressure is not zero, $E_f$ is the enthalpy of formation). Therefore,} the Pareto front 1 is the set of the most optimal solutions in terms of stability and thermoelectric efficiency. The post-processing phase focused on the front 1, and after an analysis of symmetries using STM4\cite{Valle2005a} and the calculation of $ZT$ by BoltzTraP, most of the metastables structures (a couple of examples shown in Fig. \ref{metastable_ab})  were discarded due to the their low symmetry and/or stability. At the end, two crystal structures were identified as the most optimal solutions. We now describe them in the following subsection.
\subsection{Post-processing of Structural, Electronic, and Thermoelectric Properties}
\begin{figure}[h]
\begin{subfigure}{0.28\textwidth}
  \includegraphics[width=\textwidth]{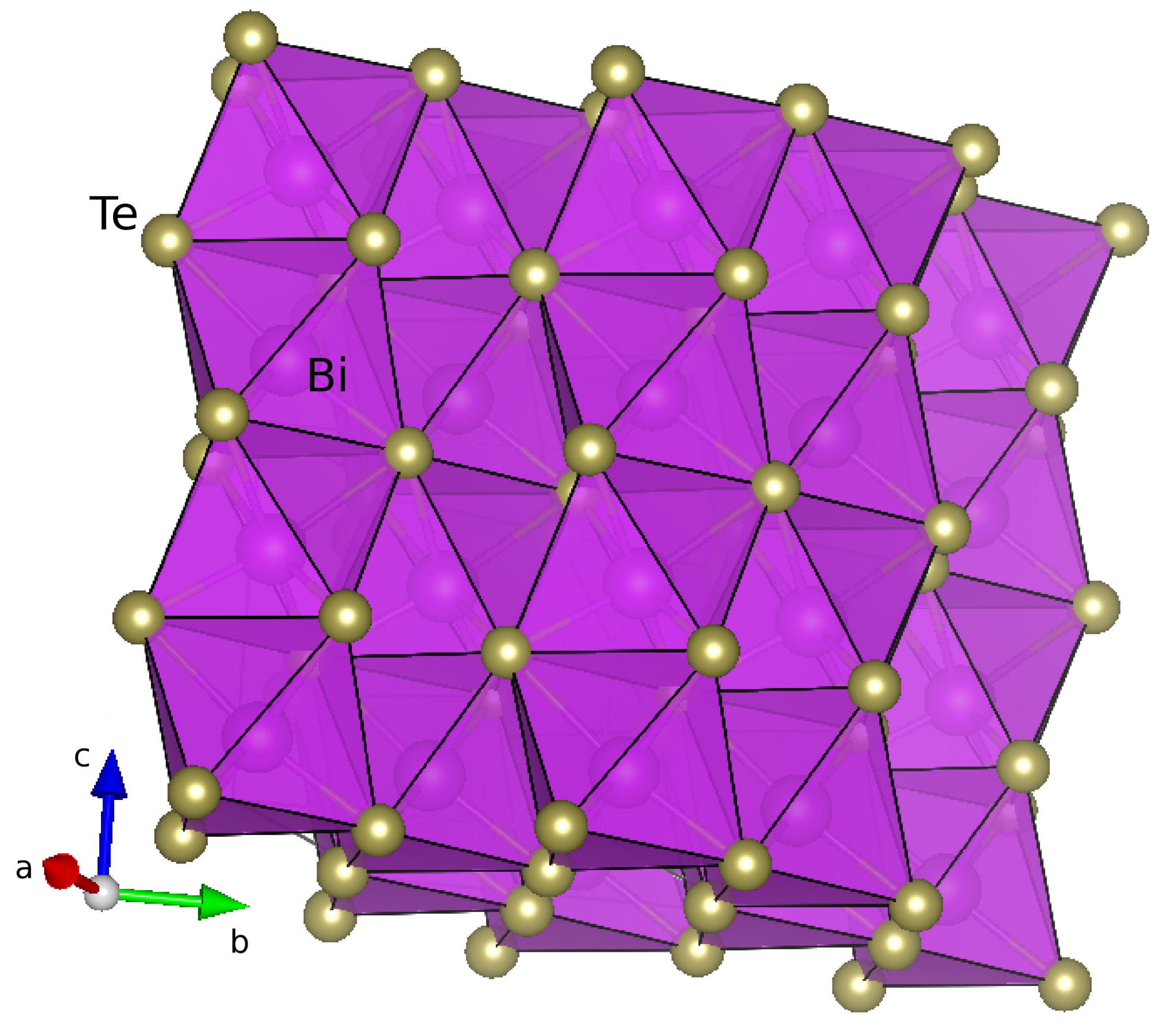}
  \caption{}
  \label{structure_a}
\end{subfigure}
\begin{subfigure}{0.3\textwidth}
        \includegraphics[width=\textwidth]{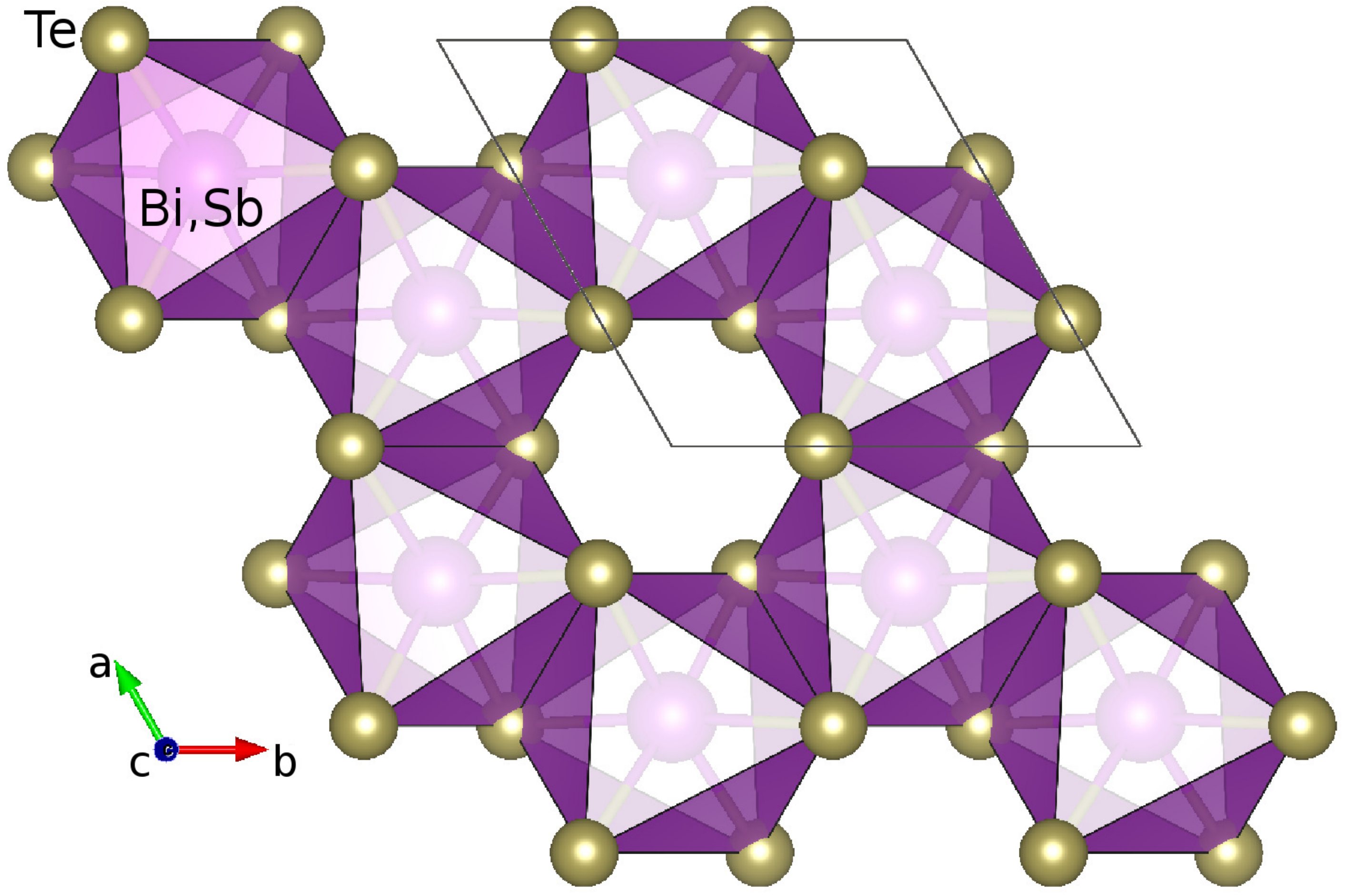}
        \caption{}
        \label{structure_b}
    \end{subfigure}
\caption{(Color online) Two views of the newly identified crystal structure in the Bi$_2$Te$_3$-Sb$_2$Te$_3$ system. The space group of this phase is $P6_3cm$ and contains two formula units/unit cell.}
\label{structure_ab}
\end{figure}

From the two phases identified with the most potential in terms of stability and thermoelectric efficiency, one is the trigonal $R\overline{3}m$  crystal structure of \textcolor{black}{Bi$_2$Te$_3$ (Sb$_2$Te$_3$)}, which is well known as a high-performance thermoelectric as well as a topological insulator. The other crystal structure shown in Fig. \ref{structure_ab}, as far as we know, has never been identified in the literature within the Bi$_2$Te$_3$-Sb$_2$Te$_3$ system. It belongs to the hexagonal space group $P6_3cm$, this group however, was predicted to lead to a topological band insulator protected by both time reversal symmetry and space group lattice symmetries\cite{Slager2013}. 
This structure type can be described as a hexagonal close packing of Te atoms, in which Bi atoms fill $\sfrac{2}{3}$ of the octahedral voids. The pattern of Bi atoms distribution is such that each Te atom has a non-planar square umbrella coordination, with the lone electron pair pointing towards the nearest empty octahedral void. One can clearly see corundum-like layers of BiTe$_6$-octahedra (edge-sharing within the layer, see Fig. \ref{structure_a}), stacked on top of each other to produce face-sharing contacts. Formally, this structure can be considered as a polytype of corundum structure, differing only in the stacking of topologically identical layers.\\

In Table I, we  compare the structural parameters of the two phases. Atomic positions and lattice parameters were further optimized as described in Section II, but using higher precision than the raw results obtained by USPEX, and also taking into account spin-orbit coupling (SOC). Brillouin zone integrations were carried out using uniform $\Gamma$-centered $9\times 9\times 2$ and $5\times5\times 5$ grids for the phases belonging to the space groups $R\overline{3}m$ (15 atoms/unit cell) and $P6_3cm$ (10 atoms/unit cell), respectively. Convergence was assumed when forces on each atom were smaller than 1 meV/\AA~ and the total energies changed by less than 1 $\mu$eV. Furthermore, our DFT calculations also suggest that the  $P6_3cm$ structures are $\sim$79 meV/atom and $\sim$65 meV/atom ($\sim$71 meV/atom and $\sim$62 meV/atom with SOC) higher in energy than the $R\overline{3}m$ structures of  Bi$_2$Te$_3$ and Sb$_2$Te$_3$, respectively. Whether or not this energy difference, $\Delta E_{str}=E_{P6_3mc}-E_{R\overline{3}m}$, is large/small enough for the realization of the  $P6_3cm$ phases, remains to be explored.\\ 

\begin{table}[h]%
\caption{\label{structure_par}%
Our calculated structural parameters at zero temperature for the two most optimal phases in the Bi$_2$Te$_3$-Sb$_2$Te$_3$ system obtained from our evolutionary simulation. Experimental values for comparison are taken from a) Ref.~[\onlinecite{Huang2008}], b) Ref.~[\onlinecite{Sokolov2007}], and c) Ref.~[\onlinecite{Anderson1974}].}
\begin{ruledtabular}
\begin{tabular}{lldddd}
   Sp. Gr.  & SOC & \multicolumn{1}{c}{\textrm{$a_0=b_0$ (\AA)}} &  \multicolumn{1}{c}{\textrm{$c_0$ (\AA)}} &  \multicolumn{1}{c}{\textrm{$V$ (\AA$^3$/f.u.)}} &  \multicolumn{1}{c}{\textrm{f.u.}} \\
   \colrule \\
   \multicolumn{6}{c}{\textrm{Bi$_2$Te$_3$}} \\
   $R\overline{3}m$  & no & 4.45 & 31.97 & 182.70& 3 \\
                           &  yes & 4.47 & 31.15 & 180.00& 3 \\
                           &      & 4.36(a) & 30.38(a) & &  \\
                           &      & 4.38(b) & 30.44(b) & &  \\
   $P6_3cm$ & no &7.48 &  7.70 &186.36 & 2 \\
          & yes &  7.47 &  7.80 &188.57 & 2 \\
   & & & & & \\
  \multicolumn{6}{c}{\textrm{Sb$_2$Te$_3$}} \\
  $R\overline{3}m$  & no & 4.34 & 31.45 & 170.70  & 3 \\
  & yes & 4.34 & 31.33& 170.17  & 3 \\
  &     & 4.27(c) & 30.47(c) & &  \\
  $P6_3cm$ & no & 7.34  & 7.85 & 183.06 & 2 \\    
     & yes & 7.35  & 7.89 & 184.30 & 2 \\    
\end{tabular}
\end{ruledtabular}
\end{table}
\begin{figure}[h]
\includegraphics[width=0.5\textwidth]{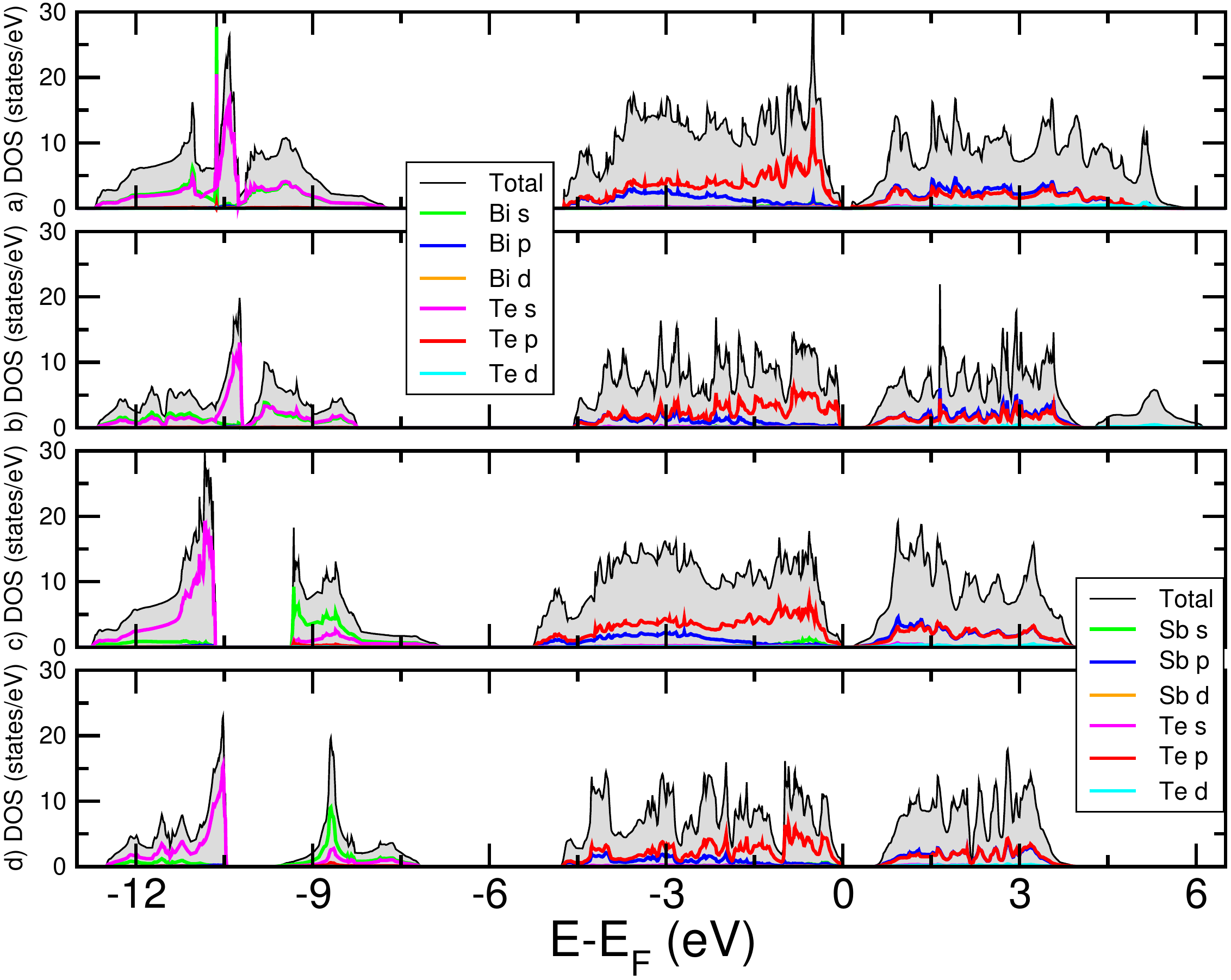}
\caption{\label{elec_dos} (Color online) Total and partial density of states (DOS) of bulk a) Bi$_2$Te$_3$ $R\overline{3}m$, b)  Bi$_2$Te$_3$ $P6_3cm$, c) Sb$_2$Te$_3$ $R\overline{3}m$, and d) Sb$_2$Te$_3$ $P6_3cm$.} 
\end{figure}

In Fig. 4, we show our calculated density of states with the inclusion of spin-orbit coupling. We notice that for both compounds the energy band gaps ($E_g$'s) increase under $P6_3cm$ symmetry from  0.14 eV to 0.30 eV, and from 0.13 eV to 0.45 eV for Bi$_2$Te$_3$ and Sb$_2$Te$_3$, respectively. The $R\overline{3}m$-$E_{g,Bi_2Te_3}$ value is in good agreement with the result obtained using WIEN2k of 0.12 eV in Ref.~[\onlinecite{RahnamayeAliabad2014157}], and with experimental values of 0.11 eV\cite{Zimmer2005} and 0.16 eV \cite{Kholer1976a,Kholer1976b}. Our computed band gap $R\overline{3}m$-$E_{g,Sb_2Te_3}$ is larger than another one also calculated with VASP of 0.09 eV\cite{sb2te3_dft_2014}, however, somewhat closer to the experimental range from 0.29 to 0.46 eV\cite{sb2te3_exp_2006,sb2te3_exp_2009}. In all scenarios, Bi, Sb and Te $s$ states lie below -6 eV in the valence band. The top of the valence and bottom  of the conduction bands are mostly dominated by Bi and Te $p$ states.\\
\begin{figure}[h]
\includegraphics[width=0.5\textwidth]{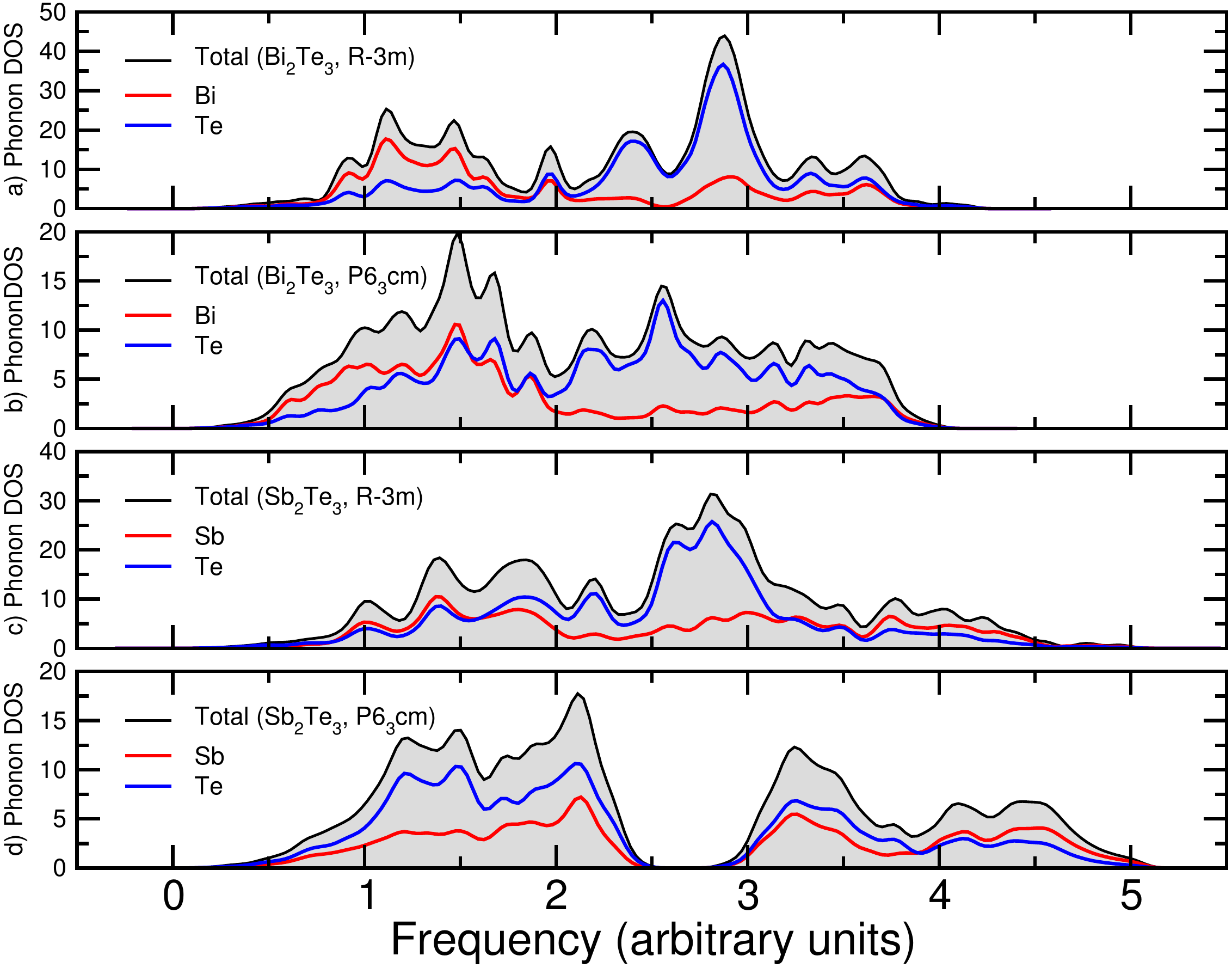}
\caption{\label{phonon_dos} (Color online) Phonon density of states of bulk a) Bi$_2$Te$_3$ $R\overline{3}m$, b)  Bi$_2$Te$_3$ $P6_3cm$, c) Sb$_2$Te$_3$ $R\overline{3}m$, and d) Sb$_2$Te$_3$ $P6_3cm$ structures at 0 K.} 
\end{figure}

To inspect the stability of the newly identified $P6_3cm$ phases, we used the finite-displacement method as implemented in the Phonopy code\cite{phonopy} for $2\times2\times2$ supercells, and calculated their phonon density of states. We then compared them to the $R\overline{3}m$ structures in Fig. \ref{phonon_dos}. These phonon spectra, in particular of  $P6_3cm$-\textcolor{black}{Bi$_2$Te$_3$ and $P6_3cm$-Sb$_2$Te$_3$},  do not show any phonon states at imaginary frequencies, which is an indication of their dynamical stability. Note, however, how the phonon densities change between the $R\overline{3}m$ and $P6_3cm$ phases. This is particularly noticeable for Sb$_2$Te$_3$.\\

\begin{figure}[h]
\includegraphics[width=0.42\textwidth]{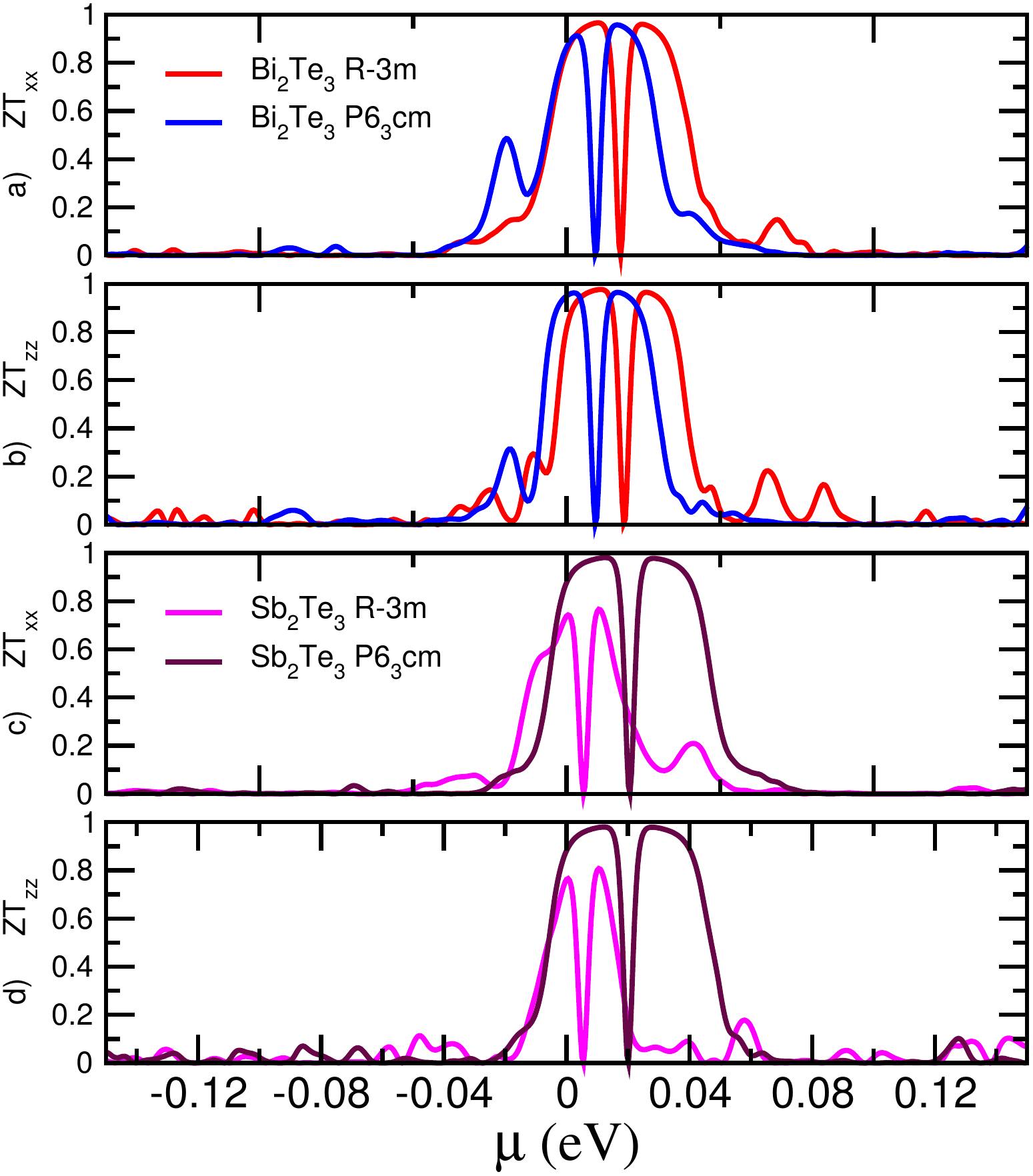}
\caption{\label{zt} (Color online) Thermoelectric figure of merit (only electrical thermal conductivity concerned) as a function of chemical potential at 300K.} 
\end{figure}

Finally, the thermoelectric efficiency for these compounds was estimated at 300 K using the BoltzTraP code with interpolated k-meshes from structural relaxations to be six times as dense. Computations were within the constant relaxation time approximation (here, $\tau$ is one unit of 10$^{-14}$ s), without taking into account the lattice thermal conductivity $\kappa_l$ and SOC. \textcolor{black}{We also should notice that, in general, the transport properties of real materials are direction-dependent, for example, the electrical conductivity is of tensorial nature\cite{Ashcroft}, $\sigma_{\alpha\beta}$. Therefore, with our method we obtain the magnitude of $ZT$ in different directions, which for isotropic materials should reduce to only one value. In the case of the Bi$_2$Te$_3$ and Sb$_2$Te$_3$ compounds, due to their $R\overline{3}m$-symmetry, $ZT$ is conventionally measured for the basal plane and along the trigonal axis, i.e., $ZT_{xx}=ZT_{yy}$ and $ZT_{zz}$. Accordingly,} the highest values of the figure of merit resulted to be $ZT_{Bi_2Te_3,zz}\sim$0.98 and $ZT_{Sb_2Te_3,zz}\sim$0.80 (see Fig. \ref{zt}). These $ZT$-maxima are moderately in agreement (overestimated) with respect to the known experimental results for the $R\overline{3}m$ phases ($ZT_{Bi_2Te_3}\sim$0.8 and $ZT_{Sb_2Te_3}\sim$0.33)\cite{Satterthwaite1957,Mehta2012}, and with the computed $ZT_{Bi_2Te_3,zz}\sim0.88$\cite{Scheidemantel2003}, but using the experimental lattice parameters and with the addition of SOC and the experimental value for $\kappa_l$. \textcolor{black}{At this point we note, how much the differences in structural parameters, Table \ref{structure_par}, and particularly in phonon properties, Fig. \ref{phonon_dos}, of the compounds can impact a material's measured $ZT$. While our first-principles method predicts somewhat comparable $ZT$ values for Bi$_2$Te$_3$ and Sb$_2$Te$_3$, without taking into account $\kappa_l$, its consideration could change remarkably the final results, as in the case of $R\overline{3}m$-Sb$_2$Te$_3$.}\\

\textcolor{black}{The calculation of $\kappa_l$ by first-principles is, however, in itself a non-straigthforward problem that is extremely constrained by implementation and computer time. ShengBTE\cite{ShengBTE} is a recent development that deals with the calculation of $\kappa_l$ based on a full iterative solution of the Boltzmann transport equation by using as main inputs sets of second- and third-order interatomic force constants (IFCs). Here, we test ShengBTE, independently of our first-principles search of efficient thermoelectric compounds, and compute the thermal conductivity of our new phase $P6_3cm$-Bi$_2$Te$_3$ and of $R\overline{3}m$-Bi$_2$Te$_3$ for reference. Following ShengBTE's prescription, we used our VASP results for structural parameters, Born effective charges and dielectric tensors. For second-order IFCs, the results of the above Phonopy calculations were directly used, while for third-order IFCs, $2\times2\times2$-supercells were built, a finite-difference approach was employed, and a cutoff radius including up to the third-nearest neighbors was enough to get satisfactorily converged values. Symmetries were also considered to generate a minimal set of displaced supercell configurations. After harmonic and anharmonic IFCs were obtained, we ran ShengBTE with $15\times15\times15$ and $12\times12\times12$ {\bf q}-point grids for the $R\overline{3}m$ and $P6_3cm$ structures, respectively. Finally, the results of the thermal conductivity computations give for $R\overline{3}m$-Bi$_2$Te$_3$ a $\kappa_{l}$=1.52 W/mK at 300 K, in comparison the experimental value for this structure is $\sim$2 W/mK\cite{Snyder2008,Chhatrasal2016}. For the $P6_3cm$-Bi$_2$Te$_3$ phase, we find $\kappa_{l}$=0.67 W/mK, and lower thermal conductivity is favorable for high-$ZT$ materials. Including $\kappa_l$, we obtain that orientationally-averaged $ZT$ of the P6$_3$cm-Bi$_2$Te$_3$ is about 2.6 times larger than $ZT$ of $R\overline{3}m$-Bi$_2$Te$_3$. Hence}, the results of $ZT$ for the  $P6_3cm$ phases, \textcolor{black}{at the level of our first-principles implementation}, suggest their potential as efficient thermoelectric materials for future and deeper investigations. The establishment of their possible topological properties is beyond the scope of this report and it will remain for future work. 
\section{Summary}
To predict crystal structures of efficient thermoelectric compounds, we have tested an evolutionary Pareto optimization search. The key feature of this approach is the simultaneous optimization of the energy and figure of merit, this strategy allows to group the results of thousands of structures in Pareto fronts with the set number 1 being the most optimal. \\

Discoveries of efficient thermoelectric compounds have been conducted either by {\it ad hoc} extensive searching or by chemical intuition. However, our evolutionary algorithm does not rely on any prior knowledge, and could be particularly useful for predicting stable and high-performance thermoelectric crystal structures. As we have shown, USPEX found the correct $R\overline{3}m$ crystal structure as the most stable structure with a large thermopower. In addition the $P6_3cm$ phases were also found by USPEX in the same calculation.\\

Of course, our approach is limited by the exclusion of the phonon part of the thermal conductivity, but its primary goal is an initial screening and ranking of structures of efficient thermoelectric materials with a desired composition. Despite these shortcomings, our method is universal and robust, \textcolor{black}{and based on a few structure generation and selection criteria, it} enables efficient structure prediction of materials with good thermoelectric efficiency {\it without the input of any experimental information}, and it can find both the stable and low-energy metastable structures in a single simulation. Above all, we hope that our work will spark further first-principles efforts to design new efficient thermoelectric materials. 

\begin{acknowledgments}
This work is supported by the Russian Science Foundation (grant 16-13-10459). Our laboratory is funded by the Project 5--100 of MIPT. Computations were performed on the Rurik supercomputer at MIPT.
\end{acknowledgments}

\bibliography{library_mari_nv}

\end{document}